\documentstyle[12pt]{article}
\newcommand{\be}{\begin{equation}}
\newcommand{\ee}{\end{equation}}
\newcommand{\ba}{\begin{eqnarray}}
\newcommand{\ea}{\end{eqnarray}}
\newcommand{\ed}{\end{document}}

\newcommand{\bfr}{\begin{flushright}}
\newcommand{\efr}{\end{flushright}}
\newcommand{\bfl}{\begin{flushleft}}
\newcommand{\efl}{\end{flushleft}}
\renewcommand{\baselinestretch}{1}
\date{}
\begin{document}

\title{SPECTRA OF QUARKONIA AT FINITE TEMPERATURE}
\author{D.U.~Matrasulov \\
Department of Heat Physics, Uzbek Academy of Sciences, \\
28 Katartal St., 700135 Tashkent, Uzbekistan \\
dmatrasu@phys.ualberta.ca \\
F.C.~Khanna\\
Department of Physics, Alberta University, Edmonton Alberta, \\
T6G 2J1 Canada and TRIUMF, 4004 Westbrook Mall Vancouver, \\
British Columbia, Canada, V6T2A3 \\
khanna@phys.ualberta.ca\\
\\
Kh.T.~Butanov and Kh.Yu. Rakhimov \\
Department of Heat Physics, Uzbek Academy of Sciences, \\
28 Katartal St., 700135 Tashkent, Uzbekistan \\
khakimjan@yahoo.com; rakhimov@infonet.uz}
\maketitle
\begin{abstract}
Finite-temperature spectra of heavy quarkonia are calculated by combining potential model and thermofield dynamics formalisms. The mass spectra of the heavy quarkonia with various quark contents are calculated. It is found that binding mass of the quarkonium decreases as temperature increases.a

Quarkonia; finite-temperature spectra; thernofield dynamics.
\end{abstract}

\section{Introduction}\label{intr}

Recent progress on creation hot and dense matter using LHC and RHIC lead to new challenges in particle physics. In particular, it has been reported recently that creation of quark-gluon plasma has been achieved in the relativistic heavy ion collisions experiments~\cite{NA50}. A key signal in the experiments on quark-gluon plasma formation is the $J/\Psi$ suppression~\cite{NA50,chai,malik,matsu} which implies that the quarkonium properties are changed under the influence of in-medium effects, such as high density and temperature. Therefore precise knowledge or qualitative estimates of these changes is of importance for such experiments.

Heavy quarkonia have been the subject of extensive theoretical and experimental research for last three decades ~\cite{eicht}-\cite{bram05}. It is well-known that the properties of mesons and nucleons change in media of high density and at finite temperature. Even the coupling constants of meson-nucleon interaction are altered as the temperature increases to the deconfining temperature. The present study is focussed on the properties of quarkonia at finite temperature~\cite{rakh00,yakh03}. Several models have been used to describe the spectra and other properties at zero temperature~\cite{mukh93}. Among others, nonrelativistic QCD~\cite{manohar,pineda,karsch}, the lattice QCD~\cite{pineda,thack,dav} and the  potential nonrelativistic QCD~\cite{kis} lead to considerable progress in understanding the nature of quark-quark and quark-antiquark interactions and predicting various properties such as decay, mass spectra, fragmentation etc. Also, a quantum mechanical approach based on potential models~\cite{eicht,rosner,eicht1} can be used to obtain quarkonium spectra for various quark contents. In this case one may use Schr\"odinger~\cite{eicht,rosner} or Bethe-Salpeter~\cite{mukh93,zhu} equations for calculation of quarkonium spectra. Despite the considerable progress made in quarkonium physics most treatments have been limited to considering zero-temperature case. There are few papers in the literature with finite-temperature treatment of quarkonium properties.

However, in medium effects on quarkonium properties should be taken into account in considering quarkonia in quark-gluon, hadronic and nuclear matter. In experiments on quark-gluon plasma formation through high energy heavy ion collisions, heavy quarkonium states are the simplest probes that allow us to test the structure of the QCD vacuum. Especially, here is of interest understanding of the behavior of heavy quark bound states in a strongly interacting media close to deconfinement temperature. Also, strict treatment of the $J/\Psi$-suppression due to color-screening should be also carried out with account of finite temperature. Few attempts have been made to use finite-temperature lattice QCD~\cite{fin}, perturbative QCD at finite temperature~\cite{arleo1} and to use color screening models~\cite{gao} in order to calculate quarkonium properties at finite temperature. Also, a hadron string model has been applied to find temperature-dependence of charmonium~\cite{chai}. In addition, the prediction of the quarkonium spectra with account of finite-temperature effects is of importance for ongoing and forthcoming experimental studies on finding of hadrons in dense and hot matter.

In this paper we present an approach for obtaining quarkonium spectra at finite temperature. Combining potential model approach and a real-time finite-temperature field theory, thermofield dynamics, we calculate the mass spectra of charmonium at finite temperature and compare it with corresponding zero-temperature ones. Finite-temperature mass spectra of the heavy quarkonia have been recently calculated on the basis of solution of the temperature-dependent Bethe-Salpeter equation~\cite{malik} and using hadron string model~\cite{chai}.

This paper is organized as follows: in the next section we give basic descritpion of the Thermofield Dynamics. In section~\ref{ftspec} thermofield dynamics formalism is applied for the calculation of the quarkonium energy spectra at finite temperature. Section~\ref{resdis} describes obtained numerical results and their discussion.

\section{Thermofield dynamics} \label{tfd}

Thermofield dynamics (TFD) is a real time operator formalism of quantum field theory at finite temperature in which any physical system can be constructed  from a temperature dependent vacuum $|0(\beta)\rangle$ which is a pure state~\cite{um,tak,das,gui}. The thermal average of any operator is equal to the expectation value between the pure vacuum state $|0(\beta)\rangle$ defined by applying Bogolyubov transformations to the usual vacuum state. TFD is based on two features. The first one is the doubling of the Fock space such that the original Fock space and its double are defined non-tilde and tilde space respectively. All operators are also doubled and the finite-temperature creation and annihilation operators are constructed by Bogolyubov transformation between tilde and non-tilde operators. This is the same procedure in writing down the vacuum state at finite temperature. Mathematically, the field operators have the following properties:
$$ (A_i  A_j\tilde ) = \tilde{A_i}\tilde{A_j}, $$
$$ (cA_i + A_j\tilde) =c^*\tilde{A_i} + \tilde{A_j}, $$
$$ (A_i^*\tilde) = (\tilde{A_i})^{\dagger}, $$
$$ (\tilde{A_i}\tilde) = A_i, $$
$$ [\tilde{A_i}  A_j] = 0. $$
It is well known~\cite{Ademir} that in an algebraic approach the doubled set of operators may be considered as a set of operators that relate to the physical observables, $O$, and a second set that are generators of symmetries, $\hat O$. The hat operators are responsible, in particular, for time development and are needed for the purpose of scattering, decay and any transitions between states. The physical observables lead to the quantities that are measured in experiment. Both for physical observables and generators of symmetry after Bogolyubov transformations leading to finite-temperature creation and annihilation operators and to a pure vacuum state only the non-tilde operators are required for getting the appropriate matrix elements.

For a given Hamiltonian $H(a, a^{\dagger})$ written in terms of annihilation and creation operators, TFD prescription implies doubling, thus constructing a new Hamiltonian $\hat H(a,a^{\dagger},\tilde a,\tilde a^{\dagger})$
\begin{equation}
\hat H(a,a^{\dagger},\tilde a,\tilde a^{\dagger}) = H(a,a^{\dagger}) -\tilde H(\tilde a,\tilde a^{\dagger}).
\label{hat}
\end{equation}
Then applying Bogolyubov transformations, which are given by
$$ a = a(\beta)cosh\theta +\tilde a^{\dagger}(\beta)sinh\theta $$
$$ a^{\dagger}= a^{\dagger}(\beta)cosh\theta +\tilde a(\beta)sinh\theta $$
$$ \tilde a = a^{\dagger}(\beta)sinh\theta +\tilde a(\beta)cosh\theta $$
$$ \tilde a^{\dagger} =a(\beta)sinh\theta +\tilde a^{\dagger}(\beta)cosh\theta $$
where
$$ \beta = \frac{\omega}{k_BT}, \;\;\;\; sinh^2\theta=\left(e^\beta-1\right)^{-1}, $$
the Hamiltonian, Eq.~(\ref{hat}) can be written in a temperature-dependent form. Physical observables to be obtained using this presciption are described in terms of non-tilde annihilation and creation operators. In the next section we apply TFD prescription for the calculation of the energy spectra of heavy quarkonia.

\section{Finite-temperature spectrum} \label{ftspec}

Within the framework of the potential model, heavy quarkonium is described by the Schr\"odinger equation which is given by
$$ i\frac{\partial\Psi}{\partial t} = H\Psi $$
where $H$ is the quarkonium Hamiltonian:
$$ H = -\frac{\Delta}{2} -\frac{\alpha_s}{r} + \lambda r + V_0, $$
with $Z=\frac{4}{3}\alpha_s$ and $\alpha_s$ being strong coupling constant.

Multiplying both sides of this equation by $r^2$ and introducing the following time scaling
$$ \tau = r^{-2}t, $$
we have
$$ i\frac{\partial\Psi}{\partial\tau} = [-r^2\frac{\Delta}{2} -\alpha_s r + \lambda r^3 + V_0r^2]\Psi. $$
Then using the substitution
$$ \Psi(r,\theta,\tau) = e^{-iE\tau}\Psi(r,\theta), $$
this equation can be written in time-independent form as:
$$ [-r^2\frac{\Delta}{2} -\alpha_s r + \lambda r^3 + V_0r^2]\Psi =E\Psi. $$
Separating angular and radial variables we get
\begin{equation}
[-\frac{1}{2}\frac{d}{dr}(r^2\frac{d}{dr})+\frac{1}{2}l(l+1) - \alpha_s r + \lambda r^3 + V_0r^2]\Psi = E\Psi
\label{schrod}
\end{equation}
Introducing annihilation and creation operators as
$$ a = \frac{1}{\sqrt{2}}\frac{d}{dr}+\frac{1}{\sqrt{2}}r, \;\;\;\;a^{\dagger} = -\frac{1}{\sqrt{2}}\frac{d}{dr}+\frac{1}{\sqrt{2}}r $$
Eq.~(\ref{schrod}) can be written in terms of these operators:
\begin{equation}
[-\frac{1}{8}(a-a^{\dagger})(a+a^{\dagger})^2(a-a^{\dagger})-\frac{\alpha_s}{\sqrt{2}}(a+a^{\dagger}) + \frac{\lambda}{2\sqrt{2}}(a+a^{\dagger})^3+\frac{1}{2}l(l+1)+\frac{1}{2}V_0\left(a+a^{\dagger}\right)^2]\Psi =E\Psi,
\label{eq}
\end{equation}
Thus, the quarkonium Hamiltonian in terms of operators $a$ and $a^{\dagger}$ can be written as
\begin{equation}
F(a,a^{\dagger})=-\frac{1}{8}(a-a^{\dagger})(a+a^{\dagger})^2(a-a^{\dagger})-\frac{\alpha_s}{\sqrt{2}}(a+a^{\dagger}) + \frac{\lambda}{2\sqrt{2}}(a+a^{\dagger})^3+\frac{1}{2}l(l+1)+\frac{1}{2}V_0\left(a+a^{\dagger}\right)^2.
\label{hamilt}
\end{equation}
%Energy eigenvalues can be found from Eq.~(\ref{hamilt}) by diagonalizing of the following matrix:
Constructing the following matrix on the harmonic oscillator basis as
\begin{eqnarray}
& E(n,n')=\langle n'\mid F(a,a^{\dagger})\mid n\rangle= \nonumber \\
& =-\frac{1}{8}\Bigg[\sqrt{n(n-1)(n-2)(n-3)}\delta_{n, n'-4}+\sqrt{(n+1)(n+2)(n+3)(n+4)}\delta_{n, n'+4}- \nonumber \\
& -(2n^2+2n+3+4l(l+1))\delta_{n, n'}\Bigg]+\frac{\lambda}{\sqrt{8}}\Bigg[\sqrt{n(n-1)(n-2)}\delta_{n, n'-3}+\sqrt{(n+1)(n+2)(n+3)}\delta_{n, n'+3}+ \nonumber \\
& +3\left(n\sqrt{n}\delta_{n, n'-1}+(n+1)\sqrt{n+1}\delta_{n, n'+1}\right)\Bigg]-\frac{\alpha_s}{\sqrt{2}}\left(\sqrt{n}\delta_{n, n'-1}+\sqrt{n+1}\delta_{n, n'+1}\right)+ \nonumber \\
& +\frac{V_0}{2}\Bigg[\sqrt{n(n-1)}\delta_{n, n'-2}+\sqrt{(n+1)(n+2)}\delta_{n, n'+2}+(2n+1)\delta_{n,n'}\Bigg].
\label{energ}
\end{eqnarray}
To find finite-temperature spectrum of the heavy quarkonium we should apply the TFD prescription to the Hamiltonian, Eq.~(\ref{hamilt}).
Then we have
\begin{equation}
\hat H = H - \tilde H,
\end{equation}
and applying Bogolyubov transformations we get
$$ H = -\frac{1}{8}\{A_1A_2A_2A_1cosh^4\theta+B_1B_2B_2B_1sinh^4\theta+[A_1A_2B_2A_1+A_1B_2A_2A_1-A_1A_2A_2B_1- $$
$$ -B_1A_2A_2A_1]cosh^3\theta sinh\theta+[B_1A_2B_2B_1+B_1B_2A_2B_1-B_1B_2B_2A_1-A_1B_2B_2B_1]cosh\theta sinh^3\theta+ $$
$$ +[A_1B_2B_2A_1-A_1A_2B_2B_1-A_1B_2A_2B_1+B_1A_2A_2B_1-B_1A_2B_2A_1-B_1B_2A_2A_1]cosh^2\theta sinh^2\theta\}- $$
$$ -\frac{\alpha_s}{\sqrt{2}}\left[A_2cosh\theta+B_2sinh\theta\right]+\frac{\lambda}{2\sqrt{2}}[A_2A_2A_2cosh^3\theta+B_2B_2B_2sinh^3\theta+ $$
$$ +(A_2A_2B_2+A_2B_2A_2+B_2A_2A_2)cosh^2\theta sinh\theta+(B_2B_2A_2+B_2A_2B_2+A_2B_2B_2)cosh\theta sinh^2\theta]- $$
$$ -\frac{1}{2}V_0\left(A_2A_2cosh^2\theta+B_2B_2sinh^2\theta+(A_2B_2+B_2A_2)cosh\theta sinh\theta\right) $$
where
$$ A_1=a(\beta)-a^{\dagger}(\beta), \;\;\;\;\;\;\; A_2=a(\beta)+a^{\dagger}(\beta), $$
$$ B_1=\tilde a(\beta)-\tilde a^{\dagger}(\beta), \;\;\;\;\;\;\; B_2=\tilde a(\beta)+\tilde a^{\dagger}(\beta). $$
Again, the (finite-temperaure) energy spectrum is obtained by diagonalization of the matrix
$$ E=\langle n`,\tilde n`\mid H(\theta,a(\beta),a^{\dagger}(\beta),\tilde a(\beta),\tilde a^{\dagger}(\beta))\mid\tilde n,n\rangle= $$
$$ =\frac{1}{8}(K-L+M)-\frac{\alpha_s}{\sqrt{2}}N+\frac{\lambda}{2\sqrt{2}}O-\frac{1}{2}V_0P, $$
where $K,L,M,N,O$ are defined as:
$$ K=4l(l+1)\delta_{n,n'}+\left(cosh^4\theta+sinh^4\theta\right)\Bigg((2n^2+2n+3)\delta_{n,n'}- $$
$$ -\sqrt{n(n-1)(n-2)(n-3)}\delta_{n,n'-4}+\sqrt{(n+1)(n+2)(n+3)(n+4)}\delta_{n,n'+4}\Bigg), $$
$$ L=\left(cosh^3\theta sinh\theta+cosh\theta sinh^3\theta\right)\Bigg(4n^2\delta_{n,n'-1}+4(n+1)^2\delta_{n,n'+1}\Bigg), $$
$$ M=2cosh^2\theta sinh^2\theta\Bigg((4n^2+4n-1)\delta_{n,n'}+ $$
$$ +\left(n(n-1)-\sqrt{n(n-1)}\right)\delta_{n,n'-2}+\Bigg((n+1)(n+2)-\sqrt{(n+1)(n+2)}\delta_{n,n'+2}\Bigg), $$
$$ N=(cosh\theta+sinh\theta)(\sqrt{n}\delta_{n,n'-1}+\sqrt{n+1}\delta_{n,n'+1}), $$
$$ O=(cosh^3\theta+sinh^3\theta)\Bigg(3n\sqrt{n}\delta_{n,n'-1}+3(n+1)\sqrt{n+1}\delta_{n,n'+1}+ $$
$$ +\sqrt{n(n-1)(n-2)}\delta_{n,n'-3}+\sqrt{(n+1)(n+2)(n+3)}\delta_{n,n'+3}\Bigg), $$
$$ P=\Bigg(\sqrt{n(n-1)}\delta_{n,n'-2}+\sqrt{(n+1)(n+2)}\delta_{n,n'+2}\Bigg)\left(cosh^2\theta+sinh^2\theta\right)+ $$
$$ +2\left(n\delta_{n,n'-1}+(n+1)\delta_{n,n'+1}\right)cosh\theta sinh\theta. $$

We can obtain finite-temperature energy eigenvalues by numerical diagonalization of this matrix.

\section{Results and discussion} \label{resdis}

In Table~\ref{tab1} we present the charmonium mass spectrum, calculated by numerical diagonalization of the matrix Eq.~(\ref{energ}) for various temperatures. As is seen from this table, the masses of the charmonium states decrease by increasing temperature. This result is in correspondence with the result of recent calculations of finite-temperature mass spectrum of charmonium~\cite{chai}, where a smooth decreasing of the charmonium mass with increasing of temperature up to the deconfinement temperature is observed.

\begin{table}[ph]
\caption{Mass spectra of charmonium (GeV) at finite temperature (GeV) in
$m_c=1.46$~GeV, $\alpha_s=0.39$ $(Z=\frac{4}{3}\alpha_s)$, $\lambda=0.2$~GeV$^2$, $V_0=-0.2$~GeV.}
\label{tab1}
\begin{center}
\begin{tabular}{ccccc}
\hline
$n$&$T=0$&$T=0.1$&$T=0.15$&$T=0.2$\\
\hline
1 (l=0)&2.481&2.447&2.025&1.966\\
2 (l=0)&2.901&2.839&2.441&2.353\\
3 (l=0)&3.812&3.515&3.401&3.175\\
4 (l=0)&4.460&4.290&4.255&4.167\\
5 (l=0)&5.277&5.264&5.147&5.098\\
1 (l=1)&2.934&2.888&2.682&2.566\\
2 (l=1)&3.382&3.290&3.027&2.964\\
3 (l=1)&4.108&3.908&3.809&3.665\\
4 (l=1)&4.586&4.484&4.451&4.255\\
5 (l=1)&5.915&5.562&5.330&5.315\\
1 (l=2)&3.799&3.750&3.487&3.350\\
2 (l=2)&4.323&4.219&4.135&3.986\\
3 (l=2)&4.571&4.523&4.428&4.256\\
4 (l=2)&5.342&5.272&5.023&4.919\\
5 (l=2)&5.999&5.884&5.753&5.628\\
\hline
\end{tabular}
\end{center}
\end{table}

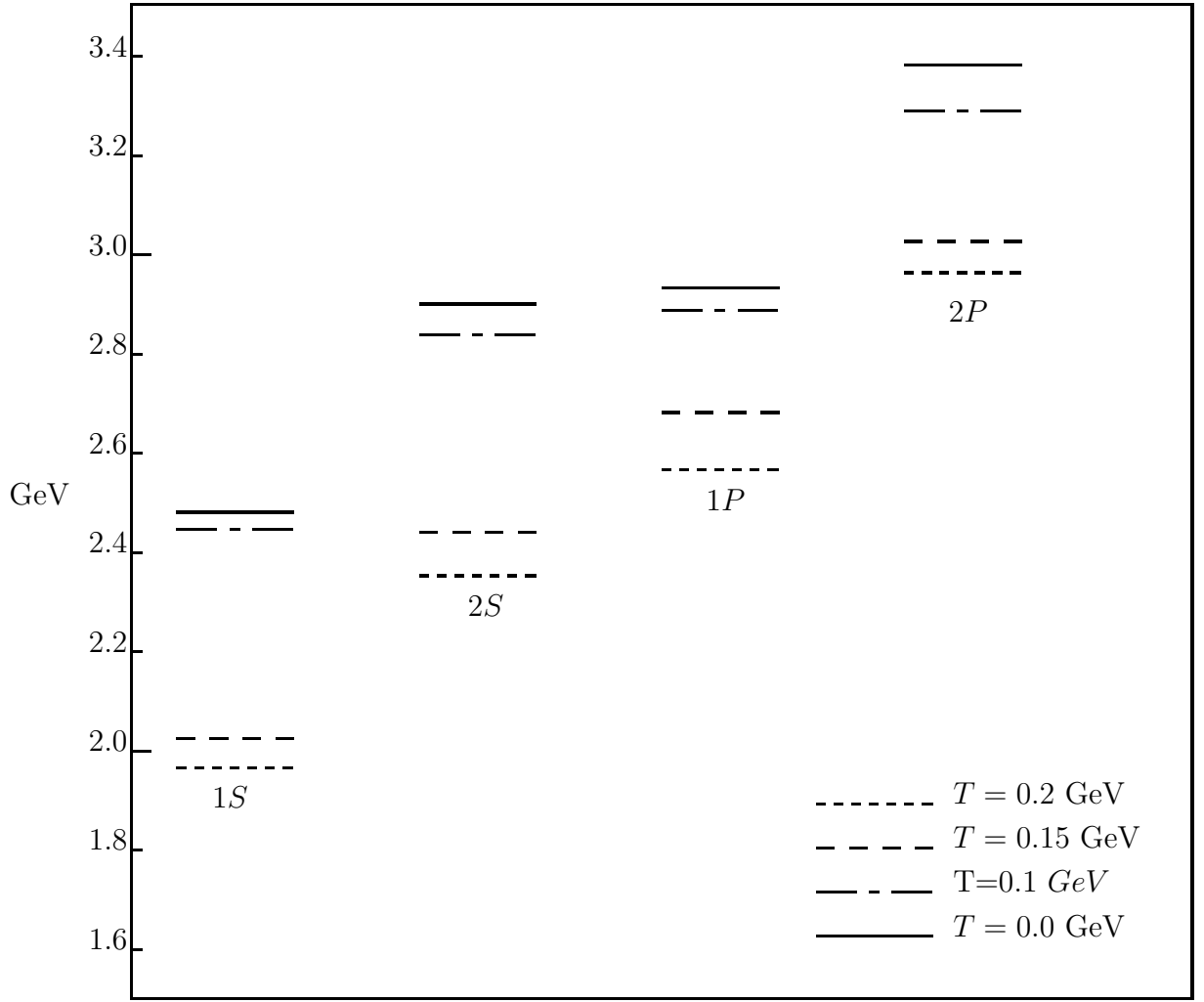
\begin{figure}[pb]
\setlength{\unitlength}{0.6mm}\thicklines
\begin{center}
\begin{picture}(200,200)
\put(30,0){\framebox(240,225)}

\put(40,52.175){\line(1,0){2}}
\put(44,52.175){\line(1,0){2}}
\put(48,52.175){\line(1,0){2}}
\put(52,52.175){\line(1,0){2}}
\put(56,52.175){\line(1,0){2}}
\put(60,52.175){\line(1,0){2}}
\put(64,52.175){\line(1,0){2.3}}

\put(40,58.8125){\line(1,0){4}}
\put(47.5,58.8125){\line(1,0){4}}
\put(55,58.8125){\line(1,0){4}}
\put(62.5,58.8125){\line(1,0){4}}

\put(40,106.2875){\line(1,0){9}}
\put(52,106.2875){\line(1,0){2.2}}
\put(57.2,106.2875){\line(1,0){9}}

\put(40,110.1125){\line(1,0){26.5}}
\put(48,43.175){$1S$}

\put(95,95.7125){\line(1,0){2}}
\put(99,95.7125){\line(1,0){2}}
\put(103,95.7125){\line(1,0){2}}
\put(107,95.7125){\line(1,0){2}}
\put(111,95.7125){\line(1,0){2}}
\put(115,95.7125){\line(1,0){2}}
\put(119,95.7125){\line(1,0){2.3}}

\put(95,105.6125){\line(1,0){4}}
\put(102.5,105.6125){\line(1,0){4}}
\put(110,105.6125){\line(1,0){4}}
\put(117.5,105.6125){\line(1,0){4}}

\put(95,150.3875){\line(1,0){9}}
\put(107,150.3875){\line(1,0){2.2}}
\put(112.2,150.3875){\line(1,0){9}}

\put(95,157.3625){\line(1,0){26.5}}
\put(106,86.7125){$2S$}

\put(150,119.675){\line(1,0){2}}
\put(154,119.675){\line(1,0){2}}
\put(158,119.675){\line(1,0){2}}
\put(162,119.675){\line(1,0){2}}
\put(166,119.675){\line(1,0){2}}
\put(170,119.675){\line(1,0){2}}
\put(174,119.675){\line(1,0){2.3}}

\put(150,132.725){\line(1,0){4}}
\put(157.5,132.725){\line(1,0){4}}
\put(165,132.725){\line(1,0){4}}
\put(172.5,132.725){\line(1,0){4}}

\put(150,155.9){\line(1,0){9}}
\put(162,155.9){\line(1,0){2.2}}
\put(167.2,155.9){\line(1,0){9}}

\put(150,161.075){\line(1,0){26.5}}
\put(160,110.675){$1P$}

\put(205,164.45){\line(1,0){2}}
\put(209,164.45){\line(1,0){2}}
\put(213,164.45){\line(1,0){2}}
\put(217,164.45){\line(1,0){2}}
\put(221,164.45){\line(1,0){2}}
\put(225,164.45){\line(1,0){2}}
\put(229,164.45){\line(1,0){2.3}}

\put(205,171.5375){\line(1,0){4}}
\put(212.5,171.5375){\line(1,0){4}}
\put(220,171.5375){\line(1,0){4}}
\put(227.5,171.5375){\line(1,0){4}}

\put(205,201.125){\line(1,0){9}}
\put(217,201.125){\line(1,0){2.2}}
\put(222.2,201.125){\line(1,0){9}}

\put(205,211.475){\line(1,0){26.5}}
\put(215,153.45){$2P$}

\put(185,44){\line(1,0){2}}
\put(189,44){\line(1,0){2}}
\put(193,44){\line(1,0){2}}
\put(197,44){\line(1,0){2}}
\put(201,44){\line(1,0){2}}
\put(205,44){\line(1,0){2}}
\put(209,44){\line(1,0){2.3}}
\put(216,44){$T=0.2$~GeV}

\put(185,34){\line(1,0){4}}
\put(192.5,34){\line(1,0){4}}
\put(200,34){\line(1,0){4}}
\put(207.5,34){\line(1,0){4}}
\put(216,34){$T=0.15$~GeV}

\put(185,24){\line(1,0){9}}
\put(197,24){\line(1,0){2.2}}
\put(202.2,24){\line(1,0){9}}
\put(216,24){T=0.1 $GeV$}
%\put(216,24){T=0.1 $\Large GeV$}

\put(185,14){\line(1,0){26.5}}
\put(216,14){$T=0.0$~GeV}

\put(2,112){GeV}
\put(30,11){\line(1,0){2}}
\put(20,11){$1.6$}
\put(30,33.5){\line(1,0){2}}
\put(20,33.5){$1.8$}
\put(30,56){\line(1,0){4}}
\put(20,56){$2.0$}
\put(30,78.5){\line(1,0){2}}
\put(20,78.5){$2.2$}
\put(30,101){\line(1,0){2}}
\put(20,101){$2.4$}
\put(30,123.5){\line(1,0){2}}
\put(20,123.5){$2.6$}
\put(30,146){\line(1,0){2}}
\put(20,146){$2.8$}
\put(30,168.5){\line(1,0){4}}
\put(20,168.5){$3.0$}
\put(30,191){\line(1,0){2}}
\put(20,191){$3.2$}
\put(30,213.5){\line(1,0){2}}
\put(20,213.5){$3.4$}
\end{picture}
\end{center}
\vspace*{8pt}
\caption{The mass spectrum of $c\overline c$ charmonium (GeV) at finite temperature (GeV). \label{fig1}}
\end{figure}

In Fig.~\ref{fig1} is the pictorial presentation of the mass spectrum for some of the charmonium states. As it is clearly seen from this figure for all presented states, the mass at $T=0.15$ GeV is considerably smaller than it is for $T=0.1$ GeV. However, the difference of masses between $T=0.2$~GeV and $T=0.15$~GeV is smaller. It should be noted that in our calculations we didn't take into account spin and color-spin effects.

Thus we have treated charmonium spectra at finite temperature using the thermofield dynamics formalism. Motivation for the treatment of quarkonium properties with account in medium effects such as finite-temperature and density comes from recent advances in quark-gluon matter creation experiments, where some states of charmonium are considered as a key signature of such matter. In addition, knowledge of the hadron spectra at finite-temperature is important for the study of hadrons in nuclear and hadronic matter. So, the obtained results can be useful in forthcoming LHC experiments on the quark-gluon plasma creation.

\section*{Acknowledgments}

The work of D.U.M., F.C.K. and Kh.Yu.R. is supported in part by NATO Collaborative Linkage Grant (PST. CLG 980300) and Kh.T.B. by a Grant of Science and Technology Center of Uzbekistan (FM 2-075).

\end{document}